\documentclass[preprint,11pt]{elsarticle}

\usepackage{amssymb}
\usepackage{amsmath}
\usepackage{wasysym}
\usepackage{tikz}

\journal{Journal of Subatomic Particles and Cosmology}

\newcommand{\blackcirc}{\raisebox{-0.5ex}{\tikz\fill (0,0) circle (0.18cm);}}
\newcommand{\whitecirc}{\raisebox{-0.5ex}{\tikz\draw (0,0) circle (0.18cm);}}
\newcommand{\bluecirc}{\raisebox{-0.5ex}{\tikz\fill[blue] (0,0) circle (0.18cm);}}
\newcommand{\yellowcirc}{\raisebox{-0.5ex}{\tikz\fill[yellow] (0,0) circle (0.18cm);}}
\begin{document}

\begin{frontmatter}

\title{The top quark in 2025---International year of Quantum Science and Technology}

\author{J. A.~Aguilar-Saavedra}

\affiliation{organization={Instituto de Física Teórica IFT-UAM/CSIC},
            addressline={c/Nicolás Cabrera 13--15}, 
            city={Madrid},
            postcode={28049}, 
            country={Spain}}

\begin{abstract}
Since its discovery at the Tevatron in 1995, the top quark has been extensively studied due to its unique properties. We discuss how the remarkable progress in top-quark physics has opened the possibility of using the top quark as a tool to test quantum mechanics at the energy frontier. After reviewing the motivations for exploring beyond quantum mechanics, we present an example: probing post-quantum correlations at the energy frontier with boosted top quarks.
\end{abstract}


\end{frontmatter}

\section{The top quark turns 30}

This year marks the 30$^\text{th}$ anniversary of the top quark discovery at the Tevatron, by the CDF and D0 Collaborations~\cite{CDF:1995wbb,D0:1995jca}. Since then, the top quark has been the subject of extensive research, owing to its properties that make it particularly interesting from both theoretical and experimental perspectives. In the following we review these aspects and the progress that has led to the first tests of quantum mechanics (QM) using the top quark.

\subsection{Experimental perspective}

From the experimental side, the top quark has several features that facilitate precise measurements of its properties. The hierarchy of Cabibbo-Kobayashi-Maskawa matrix elements $V_{td} \simeq 0.009$, $V_{ts} \simeq 0.04$, $V_{tb} \simeq 1$~\cite{ParticleDataGroup:2024cfk}, makes $t \to bW$ the only relevant decay channel, with the next one $t \to sW$ having a branching ratio around $10^{-3}$. Flavour-changing neutral decays, which are sizeable for other particles, are suppressed by a factor of $10^{-9}$~\cite{Eilam:1990zc,Aguilar-Saavedra:2002lwv} because of the Glashow-Iliopoulos-Maiani mechanism~\cite{Glashow:1970gm}. Consequently, experimental analyses are not hindered by the need to study many independent decay modes.

In addition, the short lifetime of the top quark, $\tau \simeq 5 \times 10^{-25}$\,s, simplifies the extraction of spin observables from 
angular distributions of decay products. For example, the expected value of top spin operators $\langle S_i \rangle$ can be determined in the decay $t \to bW \to b \ell \nu$ from the angular distribution of the charged lepton $\ell$ in the top-quark rest frame,
\begin{equation}
\frac{1}{\Gamma} \frac{d\Gamma}{d\Omega} = \frac{1}{4\pi} \left\{ 1 + 2 \alpha [\langle S_1 \rangle \sin\theta \cos\phi + \langle S_2 \rangle \sin\theta \sin\phi + \langle S_3 \rangle \cos\theta] \right\} \,,
\label{ec:topdist}
\end{equation}
with $\hat n = (\sin \theta \cos \phi, \sin \theta \sin \phi, \cos \theta)$ the direction of the charged lepton momentum, and $\alpha \simeq 1$ a constant dubbed as `spin analysing power'. Likewise, expected values of spin operators for the daughter $W$ boson can be measured. Furthermore, expected values of pairs $\langle S_i S_j \rangle$---the so-called spin correlations, where the two operators correspond to different particles---can be determined from higher-dimensional angular distributions.

Finally, the large top production cross sections, either in pairs or singly, allows a good statistical precision to perform the measurements.

\subsection{Theoretical perspective}

The top quark plays a central role in the hierarchy problem~\cite{Susskind:1978ms}, since its loop corrections provide the dominant quadratic contribution to the Higgs vacuum expectation value, $\delta_t v^2 \sim |y_t|^2 \Lambda^2/(16\pi^2)$,  with $\Lambda$ the ultraviolet cutoff and $y_t$ the top-quark Yukawa coupling. This has motivated the idea that the top quark may also be linked to possible solutions of this problem. In supersymmetry for example, loop corrections from top squarks cancel the quadratic divergence, $\delta_{\tilde t} v^2 \sim - |y_t|^2 \Lambda^2/(16\pi^2)$~\cite{Dimopoulos:1981zb}. Alternatively, models of dynamical electroweak symmetry breaking propose that a $t \bar t$ condensate could itself trigger electroweak symmetry breaking~\cite{Contino:2006qr}. These ideas have provided theoretical motivation to study the top quark as a possible window to new physics. However, these arguments inevitably involve a degree of {\em wishful thinking}. In particular, the characteristic mass scales suggested by such models have already been probed at the Large Hadron Collider (LHC), with no evidence for new states.

Another frequently cited motivation for top-quark studies is that, as the heaviest known particle, the top quark might exhibit enhanced sensitivity to new physics. A very simple example is a new charge-$2/3$ quark singlet $T$: its mixing is naturally expected to be larger with the third generation~\cite{Aguilar-Saavedra:2013wba}. However, these considerations involve wishful thinking as well. Even if the deviations from the standard model (SM) in top observables may be larger than for lighter fermions, the achievable experimental precision is definitely lower than in kaon and $B$ physics. Therefore, it is not guaranteed that new physics, if present, should manifest first in the top-quark sector.

The role of wishful thinking in traditional arguments advocating for the relevance of top-quark physics is not a serious drawback. Rather, it reflects the spirit that has often driven theoretical progress in particle physics. Nevertheless, it is important to acknowledge this element, in order to place in the proper context the relevance of top-quark studies for probing QM at the energy frontier.

\subsection{Three decades of studies}

A  remarkable interplay between theory and experiment has fostered the experimental investigation of top-quark properties, among them:
\begin{itemize}
\item[--] $W$ spin observables. The sensitivity of $W$ boson helicity in $t \to bW$ to physics beyond the SM was emphasised already before the top-quark discovery~\cite{Kane:1991bg}. The first model-independent determinations, requiring moderate statistics, were achieved in top pair production at the Tevatron~\cite{CDF:2008eba}. Later it was pointed out that eight independent observables fully characterise the spin density matrix of a spin-1 particle~\cite{Aguilar-Saavedra:2015yza}. Most of them have been measured in single-top production at the LHC~\cite{ATLAS:2017ygi}.
\item[--] Top polarisation. The relevance of top-spin measurements was also recognised prior to the discovery~\cite{Kane:1991bg}. The first measurements in $t \bar t$ production at the LHC~\cite{ATLAS:2013gil} focused on the top-quark helicity. For single-top production, it was noted that three independent polarisation components can be probed~\cite{Aguilar-Saavedra:2014eqa}. These measurements have since been carried out~\cite{ATLAS:2022vym}.
\item[--] Top--antitop spin correlations. It was soon realised that $t \bar t$ pairs exhibit significant helicity correlations~\cite{Stelzer:1995gc}, probed at the Tevatron shortly after~\cite{D0:2000kns}. Subsequent theoretical work extended the theoretical predictions to include correlations along orthogonal directions~\cite{Bernreuther:2015yna}, and the complete set of correlation coefficients has now been measured at the LHC~\cite{CMS:2019nrx}.
\end{itemize}
These advances in the determination of top-quark spin properties have paved the way to probing quantum correlations between the spins of the top quark and the antiquark~\cite{Afik:2020onf}, with a first measurement~\cite{ATLAS:2023fsd} that opens the way for many future studies.

\section{The top quark and quantum science}

The remarkable progress in the experimental study of the top quark has turned it into a tool to test quantum
 foundations, a possibility that was unconceivable upon its discovery. Here we review the motivation for QM tests at high energies to later focus on quantum correlations. In particular, we discuss how the top quark may be used to test post-quantum correlations.
 
\subsection{Motivation}

Tests of QM at the energy frontier bear the same importance as, for example, tests of general relativity at cosmological scales: the fact that we regard a theory as an accurate description of Nature is no reason to stop testing it across all scales. Indeed, major breakthroughs arise precisely when Nature behaves differently from what we expect from our most established theories.
 
 There are several aspects of QM that depart from classical theories, namely quantum entanglement (including Bell inequalities~\cite{Bell:1964kc}), identical particles, and contextuality. Those quantum properties can be tested at high energies in a variety of processes involving not only top quarks but also $W$, $Z$ and Higgs bosons. In the traditional approach, investigations have focused on the search for physics beyond the SM---assuming the SM is correct and looking for deviations from its theoretical predictions. Recently, attention has also turned to testing QM at the energy frontier---assuming that QM, as embedded within the SM, is correct. As previously mentioned, some degree of wishful thinking is involved, both when searching beyond the SM or beyond QM.
  
The main drawback for searches beyond QM is the absence of a compelling alternative. For example, a possible generalisation involves the inclusion of non-linear terms into the Schr\"odinger equation~\cite{Weinberg:1989us} but this was shown to lead to the possibility of superluminal communication~\cite{Gisin:1990dol}, forbidden by special relativity. But, on the other hand, at present there are no compelling alternatives to the SM either: the LHC has already explored the favoured parameter space of SM extensions addressing the hierarchy problem, without finding evidence for new particles. The absence of any new signals, neither the expected ones nor any unexpected, has in turn fueled the rise of machine-learning methods, to perform searches without assuming a specific alternative.

Besides testing the limits and validity of QM, another important point regards its interpretation. The Copenhagen interpretation of measurement (quantum state collapse) is not completely satisfactory, and indeed has been regarded by some authors as a mathematical tool rather than an ultimate description of reality. 
There are niche alternative proposals, such as Bohmian mechanics~\cite{Bohm:1951xw}, which invokes nonlocal hidden variables and thus evades the constraints imposed by Bell inequalities. However, unlike standard QM, these alternatives have not been embedded into a full quantum field theory capable of describing particle creation and annihilation. In this respect, collider processes provide new ingredients beyond traditional tests of QM, and may eventually enable novel tests of such alternative theories.

\subsection{Quantum correlations for a bipartite system}

There are several levels of quantum correlations that characterise a bipartite system. Ordered from weaker to stronger, they are:
\begin{itemize}
\item[--] Quantum discord~\cite{Ollivier:2001fdq}: non-classical correlations that may exist even in separable states (see below);
\item[--] Quantum entanglement: the two subsystems are not separable, in the sense of factorisation in Hilbert space;
\item[--] Quantum steering~\cite{Wiseman:2007hyt}: the choice of measurement performed in one party influences the quantum state of the other one;
\item[--] Bell nonlocality~\cite{Bell:1964kc}: correlations that cannot be explained by local hidden-variable (LHV) theories.
\end{itemize}
For the first three, we explicitly write the qualifier {\em quantum}  to emphasise that when identifying these types of correlations one is not addressing whether the system is quantum or not---which requires stronger correlations and cannot be addressed at those levels. In particular, some confusion is present among non-experts concerning quantum entanglement versus Bell nonlocality, which warrants clarification.

Quantum entanglement for a bipartite system, composed by two subsystems A and B, concerns the separability of the states. In the product Hilbert space $\mathcal{H}_\text{A} \otimes \mathcal{H}_\text{B}$, a pure state is said to be separable if it can be written in the form
\begin{equation}
|\psi\rangle = |a\rangle_\text{A} \otimes |b\rangle_\text{B} \,,\quad |a\rangle_\text{A} \in \mathcal{H}_\text{A} \,,  |b\rangle_\text{B} \in \mathcal{H}_\text{B} \,.
\end{equation}
Otherwise the state is said to be {\em entangled}. Mixed states are said to be separable if the density operator can be written as a sum
\begin{equation}
\rho = \sum_n p_n \rho_n^\text{A} \otimes \rho_n^\text{B} \,,
\end{equation}
with $\rho_n^\text{A}$ and $\rho_n^\text{B}$ density operators in $\mathcal{H}_\text{A}$ and $\mathcal{H}_\text{B}$, respectively, and $p_n > 0$. 

Bell nonlocality for a pair of spin-$1/2$ systems, for example a $t \bar t$ pair, is usually probed with the Clauser-Horne-Shimony-Holt (CHSH) inequalities~\cite{Clauser:1969ny}. Two experimenters, Alice  and Bob perform a series of independent measurements on their respective systems A and B: Alice measures observables $A$, $A'$ and Bob measures observables $B$, $B'$, all normalised to have eigenvalues $-1$ and $1$. Both can freely choose which observable $A/A'$, $B/B'$ to measure each time. Then, classically,
\begin{equation}
| \langle AB \rangle - \langle AB^\prime \rangle + \langle A^\prime B \rangle + \langle A^\prime B^\prime \rangle | \; \leq 2 \,.
\label{ec:CHSH}
\end{equation}
The angle brackets indicate the average over a set of measurements by Alice and Bob. In the present case that the observables measured are spins, these averages are precisely spin correlations. If the bipartite system is in some state such that the l.h.s. of (\ref{ec:CHSH}) is found to be larger than two, this implies that the correlations cannot be explained by a LHV theory.

Thus, we clearly see that separability and nonlocality are conceptually distinct notions: while the former concerns whether a state can be written as a convex mixture of product states, nonlocality refers to the impossibility of reproducing correlations with a LHV model. Moreover, it has been shown that not all entangled states exhibit nonlocality~\cite{Werner:1989zz}, therefore the hierarchy 
\begin{equation}
\text{Bell nonlocal} \subset \text{entangled}
\end{equation}
is strict. Mixed entangled states (known as Werner states) can be constructed that admit a LHV model for all projective measurements. The hierarchy has also been shown to be strict for quantum steering~\cite{Wiseman:2007hyt}.

On a different level, when addressing quantum entanglement is not necessary that the Hilbert spaces $\mathcal{H}_\text{A}$ and $\mathcal{H}_\text{B}$ describe degrees of freedom of different particles. For example, in top-quark decays there is quantum entanglement between the spins of the decay products $W$, $b$ and the orbital angular momentum (o.a.m.)~\cite{Aguilar-Saavedra:2024vpd}. While it is correct to speak about entanglement between the $W$ spin and the o.a.m, for example, it it not so meaningful to speak about Bell nonlocality.

\subsection{Quantum correlations in $t \bar t$ production}

Top pairs are produced at the LHC with strong spin correlation in certain kinematical regions.
The first measurement of quantum entanglement between the top-quark spins was performed by the ATLAS Collaboration near the production threshold~\cite{ATLAS:2023fsd}, followed by a similar measurement by CMS~\cite{CMS:2024pts}. A second region where the spin correlation is significant is for boosted central top quarks, namely when they are highly energetic and the production angle (in the centre-of-mass frame) is near $\pi/2$. A measurement of quantum entanglement in this kinematical region has also been performed~\cite{CMS:2024zkc}. 

In addition to quantum entanglement, CHSH inequalities  can be tested in the boosted central region~\cite{Fabbrichesi:2021npl}. However, it is clear that determining the l.h.s. of (\ref{ec:CHSH}) via spin correlations does not amount to probing nonlocality: in top quark decays the spin is not directly measured, nor the {\em direction} for spin measurement can be freely chosen. As mentioned in the previous section, the quantities extracted from data are expected values of spin operators, c.f. (\ref{ec:topdist}). Still, that is quite an interesting measurement to be performed, as we show in the following.

\subsection{Beyond quantum mechanics}

Quantum mechanics may violate the CHSH inequality (\ref{ec:CHSH}) for certain states and choices of observables $A$, $A'$, $B$, $B'$. On the other hand, it has also been shown that QM satisfies the so-called Tsirelson bound~\cite{Cirelson:1980ry}
\begin{equation}
| \langle AB \rangle - \langle AB^\prime \rangle + \langle A^\prime B \rangle + \langle A^\prime B^\prime \rangle | \; \leq 2 \sqrt{2} \,.
\label{ec:tsir}
\end{equation}
This limit is saturated for a pair of spin-$1/2$ particles in spin-singlet and spin-triplet states. Interestingly, top pairs in the boosted central region are produced in a nearly spin-triplet state. Therefore, the determination of the l.h.s. of (\ref{ec:tsir}) can be used to probe {\em post-quantum} correlations, that is, correlations stronger than those predicted by QM.

At this point, one may ask whether theories beyond QM could possibly exhibit stronger correlations, thereby violating the Tsirelson bound. In order to answer this question, the setup of correlation experiments can be generalised to Popescu-Rohrlich boxes~\cite{Popescu:1994kjy}. In this framework, Alice and Bob have one box each; the boxes have two buttons, say black and white, and two lights, say blue and yellow, which flash depending on the input from both boxes.\footnote{Pressing either button stands for Alice (Bob) choosing whether to measure $A$ or $A'$ ($B$ or $B'$). The two lights stand for obtaining $1$ or $-1$ as result of the measurement.}  Alice and Bob are well separated in space, they press either button at will, and record whether the blue or yellow lights flash when pressing buttons on each side.

Special relativity forbids the exchange of information faster than the speed of light (no-signalling principle); therefore, the probability for one party obtaining a given result, blue or yellow, must not depend on the button the other party chooses to press.\footnote{Assume for example that when Alice presses the black button the probability of Bob obtaining blue is 60\% (independently of the button he presses), while if Alice presses white, the probability of Bob obtaining blue decreases to 40\%. Then, Alice could use this feature to send messages in a Morse-code-like fashion at superluminal speed.} One can devise a set of rules that are consistent with this constraint:
\begin{enumerate}
\item If either Alice or Bob press black, their outputs are equal.
\item If both Alice and Bob press white, their outputs are the opposite.
\item The possible results fulfilling the above two rules have equal probability. 
\end{enumerate}
The probabilities for the different experiment outcomes, depending on the buttons that Alice and Bob choose to press, are collected in Table~\ref{tab:PR}. It is easily verified that this set of correlations obeys the no-signalling principle.

\begin{table}[htb]
\begin{center}
\begin{tabular}{|c|c|c|c|c|}
\hline
A $\backslash$ B & $\bluecirc\,\bluecirc$ & $\bluecirc\,\yellowcirc$ & $\yellowcirc \, \bluecirc$ & $\yellowcirc \, \yellowcirc$ \\
\hline
$\blackcirc \, \blackcirc$ & 0.5 & 0 & 0 & 0.5 \\
\hline
$\blackcirc \, \whitecirc$ & 0.5 & 0 & 0 & 0.5 \\
\hline
$\whitecirc \, \blackcirc$ & 0.5 & 0 & 0 & 0.5 \\
\hline
$\whitecirc \, \whitecirc$ & 0 & 0.5 & 0.5 & 0 \\
\hline
\end{tabular}
\end{center}
\caption{Probabilities of the different outcomes for the Alice--Bob boxes. Rows correspond to the combinations of buttons pressed by Alice and Bob, and columns to their recorded outcomes. The entries give the probability of obtaining each outcome for the selected input combination.}
\label{tab:PR}
\end{table}

If we make the correspondence $A,B' \leftrightarrow$ white, $A',B \leftrightarrow$ black; $+1 \leftrightarrow$ yellow, $-1 \leftrightarrow$ blue, then
\begin{equation}
\langle AB \rangle = - \langle AB^\prime \rangle = \langle A^\prime B \rangle = \langle A^\prime B^\prime \rangle = 1 \,,
\end{equation}
therefore the l.h.s. of (\ref{ec:tsir}) equals 4, larger than the maximum allowed by QM and still consistent with special relativity.
Thus, the answer to our previous question is affirmative. Theories beyond QM may indeed exhibit larger correlations, and limits on those theories could be obtained by measuring spin correlations in boosted, central production of top quark pairs.

\section{Final remarks}

There remain many open questions in QM, and high-energy colliders may provide novel insights into some of them. The first entanglement measurements in top pair production represent only an initial step, opening the way to more ambitious studies such as tests of Bell inequalities at the energy frontier. Such measurements would not only strengthen our understanding of quantum correlations in a new regime, but could also probe post-quantum correlations beyond the Tsirelson bound.

In addition, high-energy colliders offer a new ingredient not present in traditional QM experiments: particle decay. This opens new avenues of investigation, such as post-decay entanglement, which concerns the quantum correlations `inherited' after particle decays. This phenomenon has never been directly tested in experiment, and the top quark provides a unique system where such studies become feasible~\cite{Aguilar-Saavedra:2023hss}. Together, these directions underline the potential of collider physics to address questions that go beyond the scope of testing the SM, reaching into the very foundations of QM.

\section*{Acknowledgements}

Dedicated to J. Bernab\'eu, who inspired me to investigate spin beyond helicity. 
This work has been supported by the Spanish Research Agency (Agencia Estatal de Investigaci\'on) through projects PID2022-142545NB-C21 and CEX2020-001007-S funded by MCIN/AEI/10.13039/501100011033.

\end{document}